\documentclass[12pt]{article}

\usepackage{amsmath,amsfonts,amssymb}
\usepackage{array}
\usepackage{authblk}
\usepackage{epsfig}
\usepackage{graphicx}
\usepackage{hyperref}
\usepackage{bm}

\newcommand*{\Z}{\mathbb{Z}}
\DeclareMathOperator{\Tr}{Tr}

\makeatletter{}
\g@addto@macro\bfseries{\boldmath}
\makeatother

\title{\vspace*{-0.2in}
{\normalsize \hfill MIT-CTP-5406\\[0.1in]}
Identifying equivalent Calabi--Yau topologies: A discrete
  challenge from math and physics for machine learning}
\author[1]{Vishnu Jejjala\thanks{{\texttt{vishnu} \textrm{at} \texttt{neo.phys.wits.ac.za}}}}
\author[2]{Washington Taylor\thanks{{\texttt{wati} \textrm{at}
      \texttt{mit.edu}}; this note is based on a talk by WT at the
  2021 Nankai Symposium on Mathematical Dialogues in celebration of
  S.\ S.\ Chern's 110th anniversary}}
\author[3]{Andrew P.\ Turner\thanks{{\texttt{turnerap} \textrm{at}
      \texttt{sas.upenn.edu}}}}
      
\affil[1]{\small Mandelstam Institute for Theoretical Physics, School of Physics, \protect\\ NITheCS, and CoE-MaSS, University of the Witwatersrand, \protect\\ 1 Jan Smuts Avenue, Johannesburg, WITS 2050, South Africa \protect\\ ${}$}

\affil[2]{\small Center for Theoretical Physics,
    Department of Physics; \protect\\ 
    NSF AI Institute for Artificial
    Intelligence and Fundamental Interactions, \protect\\
    Massachusetts Institute of Technology, \protect\\
    77 Massachusetts Avenue,
    Cambridge, MA 02139, USA\protect\\ ${}$}

\affil[3]{\small Department of Physics and Astronomy, University of Pennsylvania, \protect\\ 209 South 33rd Street, Philadelphia, PA 19104, USA}

\begin{document}

\maketitle

\begin{quotation}
\noindent {\bf Abstract:}
We review briefly the characteristic topological data of Calabi--Yau
threefolds and focus on the question of when two threefolds are
equivalent through related topological data.
This provides an interesting test case for machine learning
methodology in discrete mathematics problems motivated by physics.

\end{quotation}
\flushbottom

\newpage
\section{Calabi--Yau threefolds}\label{sec:intro}

Calabi--Yau threefolds are a class of geometric spaces that have
been studied intensively for several decades by mathematicians and
physicists in the context of superstring compactification~\cite{chsw, Hubsch-book, He-book}.
The key features of these real 6-dimensional manifolds are (1) that they
admit a Ricci-flat K{\"a}hler metric, and thus solve the vacuum Einstein
equations $R_{\mu \nu} = 0$, and (2) they are K{\"a}hler manifolds and
thus have a complex structure compatible with supersymmetry.  This
makes them useful as geometries for compactifying superstring theory
from ten dimensions to the physical four dimensions of space-time.
Mathematically, the key feature of a Calabi--Yau variety is the
vanishing of the canonical class $K = 0$ (up to torsion).

Despite extensive investigations on these geometries, a fundamental
open question remains: \emph{Is the number of topologically distinct Calabi--Yau threefolds finite or infinite?}~\cite{Yau-66}

Topologically, a theorem of Wall~\cite{Wall} states that a Calabi--Yau threefold
$X$
is uniquely determined by:
\begin{itemize}
\item
$h^{1,1}, h^{2,1}\in \Z_+$:  Hodge numbers;

\item
$C_{i j k} \in \Z, \; i, j, k \in \{ 1, \dotsc, h^{1,1}\}$:
  triple intersection numbers between divisors (complex codimension-one subvarieties)
  in the linear space
$H^{1,1}(X, \Z)$
(where $h^{1, 1}(X) = \operatorname{dim} H^{1, 1}(X,\Z)$);

\item
$p_1 (T_X) = 2 c_2 (T_X)$: The first Pontryagin class, a vector of $h^{1,1}$ integers.
\end{itemize}

Thus, the topological data of a Calabi--Yau threefold is characterized
by a finite list of integer data.  Fundamental questions to which the
answers are not yet known are:
\begin{enumerate}
\item[(1)]
Which sets of integer data are allowed?

\item[(2)]
Which sets of integer data are equivalent under a basis change for $H^{1,1}$
(and hence describe the same geometry)?
\end{enumerate}

We are focused here on the second of these questions.

Over the years, many large classes of Calabi--Yau threefolds have been
constructed.  The largest is the set of toric hypersurface Calabi--Yau
threefolds realized through the Batyrev construction~\cite{Batyrev}
as
 hypersurfaces
in toric fourfolds; Kreuzer and Skarke have compiled a comprehensive list
of 473,\,800,\,776 reflexive 4D polytopes that provide such
constructions~\cite{KS-classification, KS-database}.  Other
constructions include complete intersection Calabi--Yau (CICY) threefolds~\cite{cdls} and generalized CICYs~\cite{Anderson-aggl}.

\begin{figure}[!t]
\centering

\includegraphics[width=6cm]{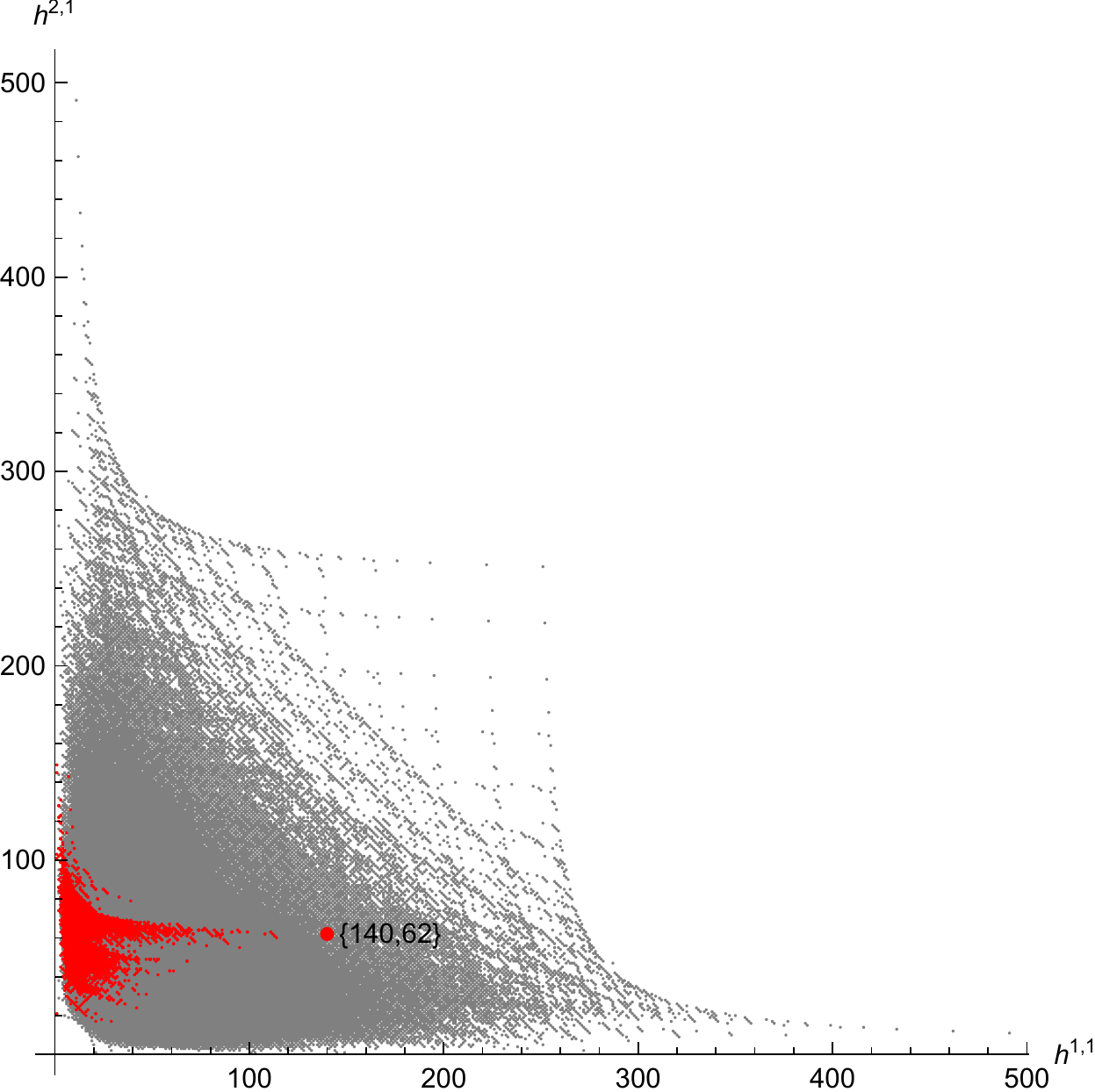}

\caption{\small Hodge numbers of the 473.8 million families
of Calabi--Yau threefolds described by Kreuzer and Skarke through
toric hypersurface constructions from 4D reflexive polytopes.  Hodge
numbers for the less than 30,000 cases without obvious elliptic fibers are
shown in red; even many of these may have more subtle elliptic
fibration structure in some phase.  Since the number of
topologically distinct elliptic Calabi--Yau threefolds is finite,
this suggests finiteness of the full Calabi--Yau landscape.}
\label{fig:Hodge}
\end{figure}

Regarding the finiteness question, recent work on elliptic Calabi--Yau
threefolds, which are particularly relevant in the context of F-theory~\cite{Vafa-F-theory, Morrison-Vafa}, has given some fairly strong evidence that the number
of distinct topological Calabi--Yau threefolds may be finite (up to birational
equivalence under flops).
An elliptic (or more generally genus one fibered) Calabi--Yau threefold
admits a fibration $\pi\colon X \to B_2$ so that over an open
subset of $B_2$ the fiber $\pi^{-1}(x)$ is an elliptic curve or genus
one algebraic curve.
It is known that the
number of topological types of elliptic Calabi--Yau threefolds is
finite~\cite{Grassi, Gross}.  Furthermore, it has been found that almost all
known Calabi--Yau threefold constructions admit an elliptic fibration in some flop
phase.  For example, 99.3\% (7837/7890) of the CICY
threefolds have an ``obvious'' elliptic or genus one fibration~\cite{aggl}, and all have at least one such fibration when $h^{1,1}
\ge 4$.  For the larger class of toric hypersurface Calabi--Yau threefolds, a
systematic analysis was carried out in~\cite{HT-all}, finding that
99.994\% of the threefolds in the Kreuzer--Skarke database have a manifest elliptic
fiber in some phase (for each polytope there are many toric varieties,
and different triangulations give different phases related under
flops).
A graph of the Hodge numbers of the Calabi--Yau threefolds in the Kreuzer--Skarke
database is shown in Figure~\ref{fig:Hodge}; the Hodge numbers of the
29,\,223 reflexive polytopes that give Calabi--Yau threefolds with no obvious elliptic
fiber are shown in red.

\section{Triple intersection numbers and topological equivalence}

As described above, an important open question is to identify a
systematic algorithm for determining when two sets of discrete data
that characterize a topological Calabi--Yau threefold are equivalent
under a basis change.  In particular, we focus on the specific part of
that problem associated with the triple intersection numbers
\begin{equation}
C_{i j k} \in \Z\,, \quad i, j, k \le N\,.
\end{equation}
These intersection numbers form a 3-index symmetric tensor so that
$C_{ijk} = C_{jik} = C_{ikj}$.

We wish to address the following mathematical question:
\paragraph{Problem:} 
\emph{Given two 3-index symmetric tensors $C, C'$ of degree $N$
associated with Calabi--Yau threefold geometries,
does there exist  an integer linear change of basis
$\Lambda_{i j} \in \operatorname{SL}(N, \Z)$ such that
\begin{equation}
C_{i j k}' =\Lambda_{i l} \Lambda_{j m} \Lambda_{k n} C_{l m n}\,?
\end{equation}}

Because the set of possible linear transformations $\Lambda$ is
infinite, there is no simple way to check the equivalence in finite
time.  In fact, there is no known algorithm to check this equivalence
in finite time, whether or not it is known that the triple
intersection numbers correspond to allowed Calabi--Yau threefold
geometries.

Some partial results in this direction have been found. In particular,
there are certain functions of the triple intersection
numbers that are invariant under a basis change; thus, if the
invariants differ between the intersection numbers $C$ and $C'$, there
clearly cannot be an equivalence. 
Some of these invariants are summarized in~\cite{Hubsch-book}; for example,
$\gcd(\{C_{i j k}\}_{1 \le i, j, k \le N})$ is clearly invariant under any
transformation $\Lambda_{i j} \in \operatorname{SL}(N,\Z)$.  Recently, some new
invariants have been identified using the methodology of limiting
mixed Hodge structures~\cite{grv}.

Because there is no known algorithm for checking this equivalence,
this is a natural candidate for the application of machine learning (ML)
methodology.
In recent years, as elaborated in some detail in the recent
text~\cite{He-book}, there has been a great deal of work on using ML
methods to solve discrete problems related to Calabi--Yau geometries,
with varying degrees of success
\cite{He:2017aed}--\cite{Berglund:2021ztg}.
On one hand, one might hope that by presenting an ML system
with a variety of data of triple intersection numbers with families of
equivalence known from construction it may be possible to find some
approximate method for checking equivalence.  More ambitiously, one
might hope to identify through this approach some systematic structure
in the triple intersection numbers such as new invariants that may
lead to an analytic solution of the problem.

Direct application of standard ML techniques to this problem does not
immediately give insight, so we turn to a related problem to
understand how machine learning can address discrete transformation
problems of this type.

\section{Machine learning and a simpler matrix problem}

As an analogue of the problem of identifying when two 3-index tensors
are equivalent under a basis change, we can consider the simpler
problem of using machine learning to figure out when two matrices
(i.e., 2-index tensors) are similarly equivalent. To relate this to
some known results, we also drop for now the condition that the
entries are integer.  We can then consider the following simpler
problem:
\paragraph{Problem:} 
\emph{Given complex matrices $A_{i j}, B_{i j}$, does there exist a unitary
transformation $\Lambda$ so that
\begin{equation}
B = \Lambda A \Lambda^\dagger\,?
\end{equation}}

There is a known set of criteria that determines when
two such matrices are equivalent in this way.
\emph{Specht's theorem}~\cite{specht1940theorie} states that
$A \sim B$ iff 
\begin{equation}
\begin{aligned}
\label{eq:conditions}
    \Tr A &= \Tr B\,, \\
    \Tr A^2 &= \Tr B^2\,, \\
    \Tr A A^\dagger &= \Tr B B^\dagger\,, \\
    &\vdots
\end{aligned}
\end{equation}
i.e.,  $\Tr W(A, A^\dagger) =\Tr W(B, B^\dagger)$ for all
words $W$ that are functions of the matrix and its adjoint.

In fact, for matrices of a given size $N$, it is sufficient to check
the agreement between these invariants only up to a certain length.
For $N = 2$, it is sufficient to check only up to words of degree 2
(i.e., 3 independent conditions). For $N = 3$, words up to degree 6
suffice (7 conditions). For general $N$ it has been proven that it is sufficient
to check up to degree $N \sqrt{2 N^2 / (N - 1) + 1/4} +
N / 2 - 2$, and it has been conjectured that the degree of sufficiency
actually scales linearly in $N$~\cite{Paz,FGG}.

Thus, we know that for matrices (2-index tensors), this is a solvable
problem even when the entries are non-integer. Thus, it is perhaps
illuminating to see how machine learning can do at learning to check
such equivalences, and we can see if a simple machine learning system
can be designed that can essentially ``discover'' Specht's theorem.

Motivated by these considerations, we have implemented some simple ML
algorithms to attempt to solve the matrix equivalence problem using the neural network functionality built in to Mathematica.

The simplest experiment is to try random \emph{real} matrices $A, B$ with a simple
multi-layer ReLU network, providing data in which half of the cases
given are equivalent matrices and the other half are inequivalent,
with no further constraints. This kind of network quickly
converges to near-100\% accuracy. Some investigation, however, shows that
this network is essentially just checking the first (linear) condition
in~(\ref{eq:conditions}), i.e., comparing the traces of the input matrices. It is not surprising that a simple ML
system can rapidly detect when two matrices satisfy this simple linear
condition.

\begin{figure}[!t]
    \centering
    
    \includegraphics[width=0.7\textwidth]{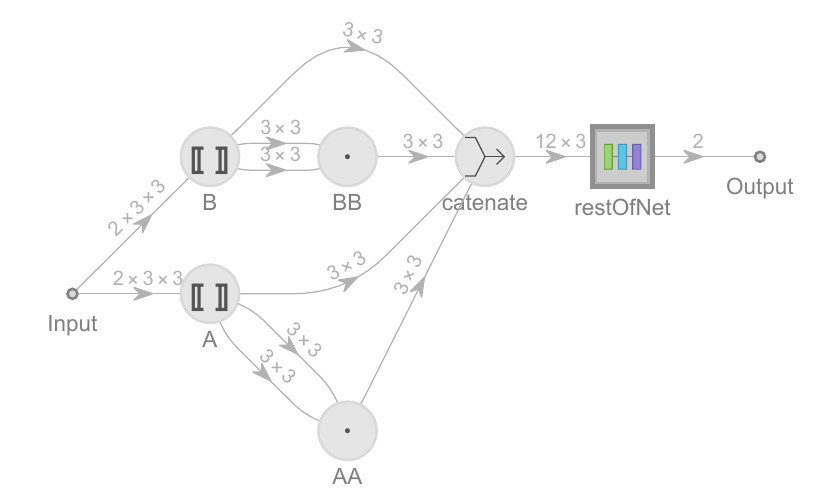}
    
    \caption{\small Network architecture for learning orthogonal equivalence of $3 \times 3$ real matrices with fixed first-order invariant. The input matrices $A$ and $B$ are fed into dot layers that compute the matrix squares, and then $A$, $A^2$, $B$, and $B^2$ are fed into \texttt{restOfNet}, which is a multi-layer ReLU network.}
    \label{fig:architecture}
\end{figure}

\begin{figure}[!t]
    \centering
    
    \includegraphics[width=0.45\textwidth]{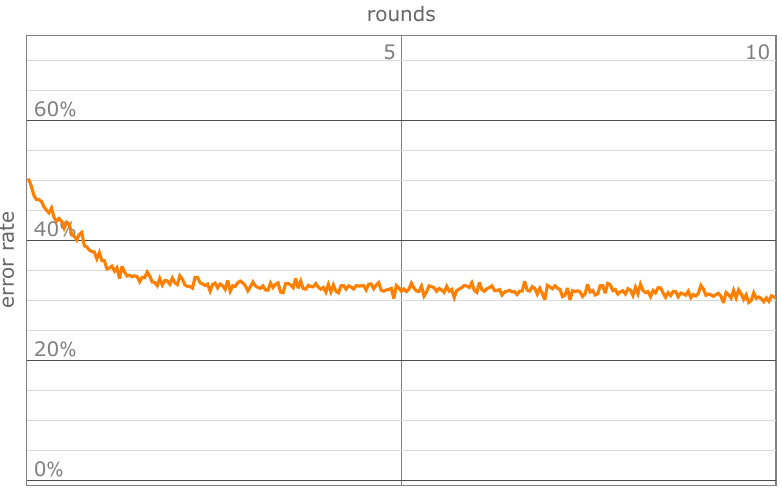}
    \includegraphics[width=0.45\textwidth]{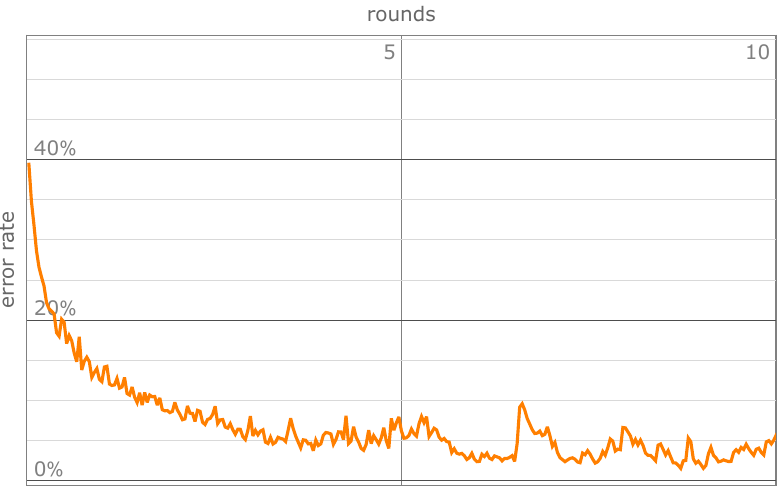}
    
    \caption{\small Error rates during training for two different network architectures learning to identify orthogonal equivalence of $3 \times 3$ real matrices with fixed first-order invariant. On the left, a multi-layer ReLU network; on the right, the architecture shown in Figure~\ref{fig:architecture}.}
    \label{fig:errors}
\end{figure}

The next challenge is to provide a data set in which we fix the
condition $\Tr A =\Tr B$ by hand for all pairs.  Using
such a dataset of $3 \times 3$ matrices, and a basic multi-layer ReLU network, with some training
the network accurately recognizes pairs that are equivalent about 75\% of
the time and pairs that are inequivalent about 60\% of the time.  This is
not a very good success rate.  By changing the network architecture to explicitly compute all products of terms in the first layer by hand (see Figure~\ref{fig:architecture}), however, the success rates
immediately increase to 99\% for equivalent pairs and 90\% for
inequivalent pairs. The error rates during training for both network architectures are shown in Figure~\ref{fig:errors}. Thus, with this change of architecture the neural
network ``discovers'' the quadratic term in Specht's theorem.
It seems, however, that higher products must similarly be included by
hand to efficiently converge on a system that knows about the
additional constraints in~(\ref{eq:conditions}). Similar results
hold for complex matrices, where we must use a slightly more
complicated version of the architecture shown in
Figure~\ref{fig:architecture} to include $\Tr A A^\dagger$ as well as
$\Tr A^2$. 

\section{Conclusions}

In this brief extended abstract we have described a basic problem in
identifying equivalence of symmetric 3-index tensors that is central
to classifying Calabi--Yau threefolds by discrete information
characterizing their topology.  We have outlined the first steps of an effort to use
machine learning to solve a simpler version of the problem with
2-index tensors where the analytic solution is known.  Several lessons
that can be taken from this are the following:
\begin{itemize}
\item Out of the box ML algorithms do not immediately yield
insight in discrete problems of this kind.

\item The fact that including product terms by hand in the ML
architecture facilitates ``discovery'' by the neural network of
nonlinear constraints suggests that networks with
different basic functional units may perform dramatically differently
on certain classes of problems; here we have an explicit understanding
through Specht's theorem of this structure.

\item It has been pointed out by, e.g.,~\cite{LTR} that
in principle a general nonlinear layer type in neural networks can
learn the multiplication function, so in principle the distinction
between these different functional units may be essentially some
factor of overhead.  It seems likely, however, that the time to learn
such things may be exponentially large so that in practice network
architecture and the form of nonlinear function used may be crucial in
getting networks to learn certain structures.
\end{itemize}

The work described here is really only a first step at using ML
methods for analyzing this problem.  It would be interesting to extend
the analysis further to reproduce higher order terms in the matrix
case and to investigate whether sufficiently sophisticated network
architectures could learn to recognize equivalent 3-index integer
tensors for the Calabi--Yau equivalence problem, and if so whether it
would be possible to extract new analytic information such as new Calabi--Yau
threefold invariants from this analysis.


\subsection*{Acknowledgments}
We would like to thank Lara Anderson, James Gray, Harold Erbin,
Tristan H\"ubsch,  Max Tegmark, and Yinan Wang
for helpful discussions.  This work arose in the context of
collaborative visits supported by a grant from the MISTI
MIT-Africa-Imperial College seed fund initiative, and was supported in
part by this grant. VJ is supported in
part by the South African Research Chairs Initiative of the National
Research Foundation, grant number 78554 and the Simons Foundation
Mathematical and Physical Sciences Targeted Grants to Institutes,
Award ID:509116.  WT is also supported by DOE grant DE-SC00012567. APT
is supported in part by the DOE (HEP) Award DE-SC0013528.
This work was also supported in part by the National Science
Foundation under Cooperative Agreement PHY-2019786.


\end{document}